# Current noise in high Tc granular superconductors under non-stationary conditions of current and magnetic field

P. Mazzetti[a], P. Tura[b], A. Masoero[b], A. Stepanescu[a]
[a]INFM-Dip. di Fisica, Politecnico di Torino, C.so Duca degli Abruzzi, 24-10129 Torino, Italy
[b]INFM-Dip. di Scienze e Tecnologie Avanzate dell'Università del Piemonte Orientale "Amedeo Avogadro", C.so Borsalino, 54-10131 Alessandria, Italy

## ABSTRACT

We present a set of experimental results concerning the power spectrum of current noise, detected on a granular high Tc superconductor submitted either to a slowly varying magnetic field or to a varying current intensity. Experiments were performed on a YBCO specimen suitably treated in order to weaken the weak links without affecting the oxygen content of grains. The weakening of the intergrain region allowed the use of very small magnetic fields and currents to induce the resistive transition of the specimen and to observe current noise. The induced noise is of the $1/f^2$ type and will be interpreted in terms of two different models. One of the model is based on the enhancement of the noise due to the clustering of the resistive transition of the weak links, produced by correlation effects related to the strong nonlinearity of their Josephson type I-V characteristics. This model has been the object of a computer simulation based on a 3D-network of Josephson-like elements and seems suitable to explain the noise produced by current variation. The second model explains the excess noise as produced by discontinuous penetration of the magnetic flux inside the intergrain region. This discontinuity is related to the field screening effect of rings made of several superconducting weak links connecting different grains, which are alternatively broken and restored by the current induced during flux variation, and seems suitable to explain the larger noise produced by a varying magnetic field.

## 1. INTRODUCTION

Current noise in granular high $T_c$ superconductors (HTSC) is in general related to the presence of the weak links. These elements are the intergrain layers which behave as shunted Josephson junction, and are characterized in most HTSC by a critical current and a critical magnetic field much smaller than the ones measured within the grains. These properties, typical of the HTSC having a granular structure, give rise to the well known two-stage transition [1] to the superconductive state, as reported in Fig. 1.

Current noise is generated when the specimen becomes resistive in all the cases where the current or the magnetic field or both these quantities exceed their critical values, at least for what concerns the weak link ensemble.

Two important cases can be considered: the case where the noise is generated in stationary conditions and the case where the current or the magnetic field or both these quantities are varying with time.

This paper is particularly dedicated to this second aspect of the noise, produced by a low frequency a.c. current or a.c. magnetic field, having sinusoidal waveform and an amplitude sufficiently small to give rise to transition within the weak link ensemble without involving the superconducting grains. To this purpose experiments were performed on a specimen of polycrystalline YBCO, suitably prepared to have weak links characterized by a much lower critical current than the crystalline grains, as described in the next section.

As observed experimentally, there is a strong difference in the noises generated in stationary and non stationary conditions. When a constant current is flowing in a HTSC made resistive by a d.c. magnetic field (or by the current itself when it is sufficiently large) a current noise is generated whose power spectrum is of the 1/f type [2,3,4,5,6], while when the current or the field are changing an additional noise component occurs whose power spectrum is of the $1/f^2$ type [7]. This component dominates the low frequency part of the power spectrum, and, in the case of the magnetic field variation, it is so large that the 1/f component is negligible and the whole spectrum appears to be of the $1/f^2$ type. As it will be shown in the following, the $1/f^2$ component is generated by a series of conductance steps related to the abrupt switching from the superconductive to the resistive state of a rather large set of weak links when the field or the current are increasing, while the opposite occurs when these quantities decrease.

The fact that each step involves the simultaneous transition of several weak links is experimentally proved by the amplitude of this noise component and it is confirmed by a computer simulation where the HTSC is represented as a 3-dimensional network of shunted Josephson junctions [8] having a gaussian distribution of critical currents [7].
The simulation shows that, when even a single junction (representing a weak link) makes a transition, the rearrangement of currents in the network produces several other transitions in order to satisfy Kirchhoff equations.
In the present paper we shall present and discuss the experimental results of current noise obtained on the cited YBCO specimen, both in the case where the current is changing and in the case where the magnetic field is changing. The results will be interpreted on the basis of the network model briefly sketched above.
The analysis of these results allows to evidence that, in the case of a changing magnetic field, in addition to the clustering effect due to the rearrangement of the current distribution following each weak link transition, there is a further clustering due to a discontinuous flux penetration which strongly enhances the $1/f^2$ component.

## 2. EXPERIMENTAL

Reported experiments were performed on a bulk polycrystalline YBCO specimen specially prepared to have a large ratio between the critical currents within the crystalline grains and within the intergrain region (weak links). This result was obtained by means of appropriate annealing in air, which had the effect of reducing the oxygen content in the intergrain region. After treatment the specimen presented a much lower critical current and a much lower critical field for what concerns the first stage of its transition to the resistive state, i.e. the stage where the grains remain superconductive and only the weak links become resistive (see Fig.1).

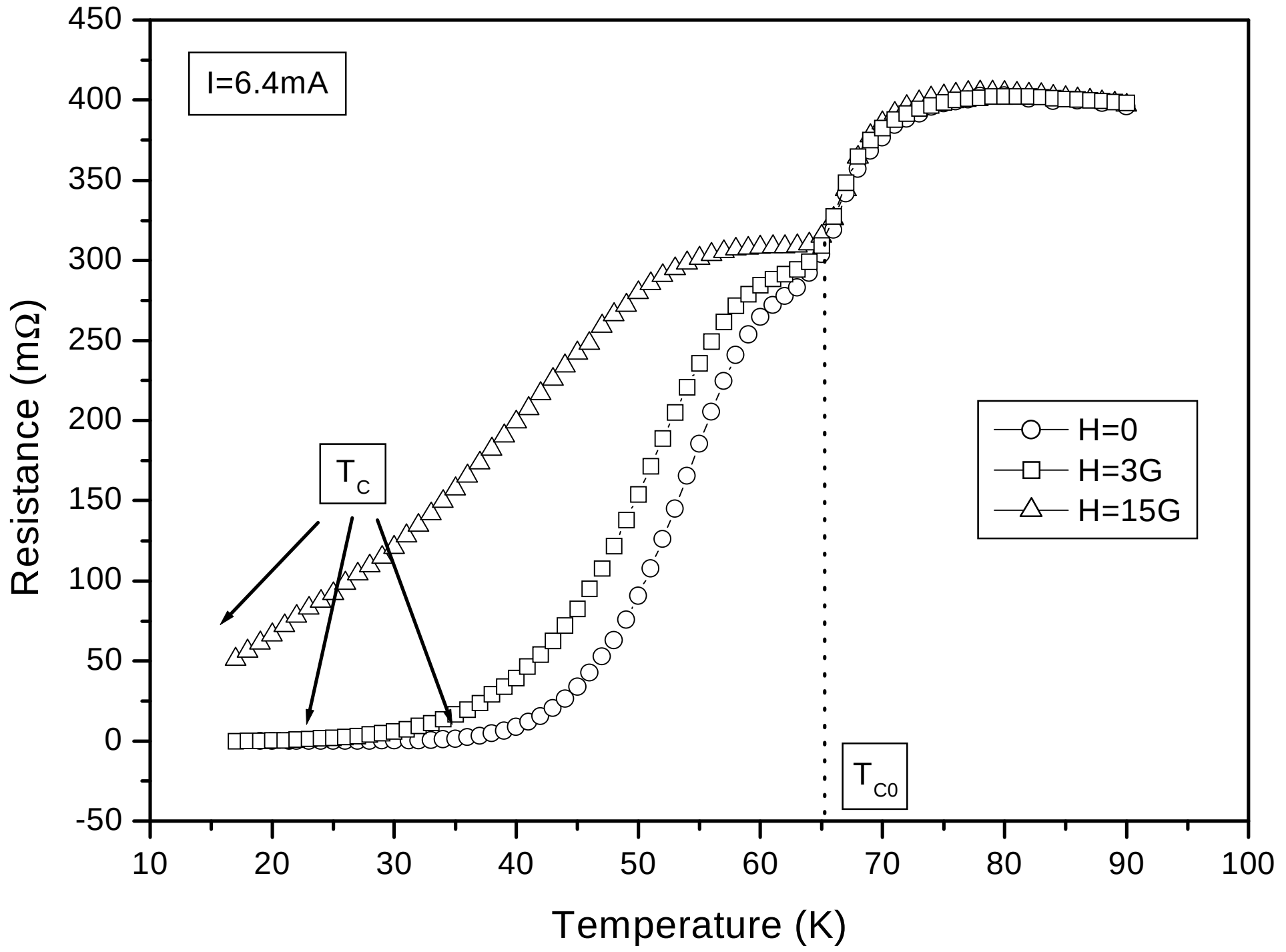

Figure 1: Experimental temperature dependence of the specimen resistance for three different values of the external magnetic field: H=0 (circles), H=3G (squares), H=15G (triangles). Current intensity was I=6.4mA. The figure clearly shows the presence of the step produced by the grain superconductive transition while the weak links are still in the resistive state, and evidences the extreme sensitivity of the specimen resistance to the magnetic field.

The electrical resistance just below $T_{C0}$ represents the saturation value corresponding to the condition where all the weak links have undergone the resistive transition and only the grains remain superconductive. High sensitivity to the magnetic field is proven by the fact that this saturation value is easily reached at temperatures much lower than $T_{C0}$ by the application of small field values, as can be observed from the results reported in Fig. 1
This situation did allow to interpret the results concerning conductivity and noise in terms of weak link transitions and to modellize the specimen as a 3-dimensional network of non linear resistors, whose nodes represented the superconducting grains and the non linear resistors, having a Josephson-like V-I characteristics, the weak links. A short description of this model and the main aspects of its results are reported in ref.[7].

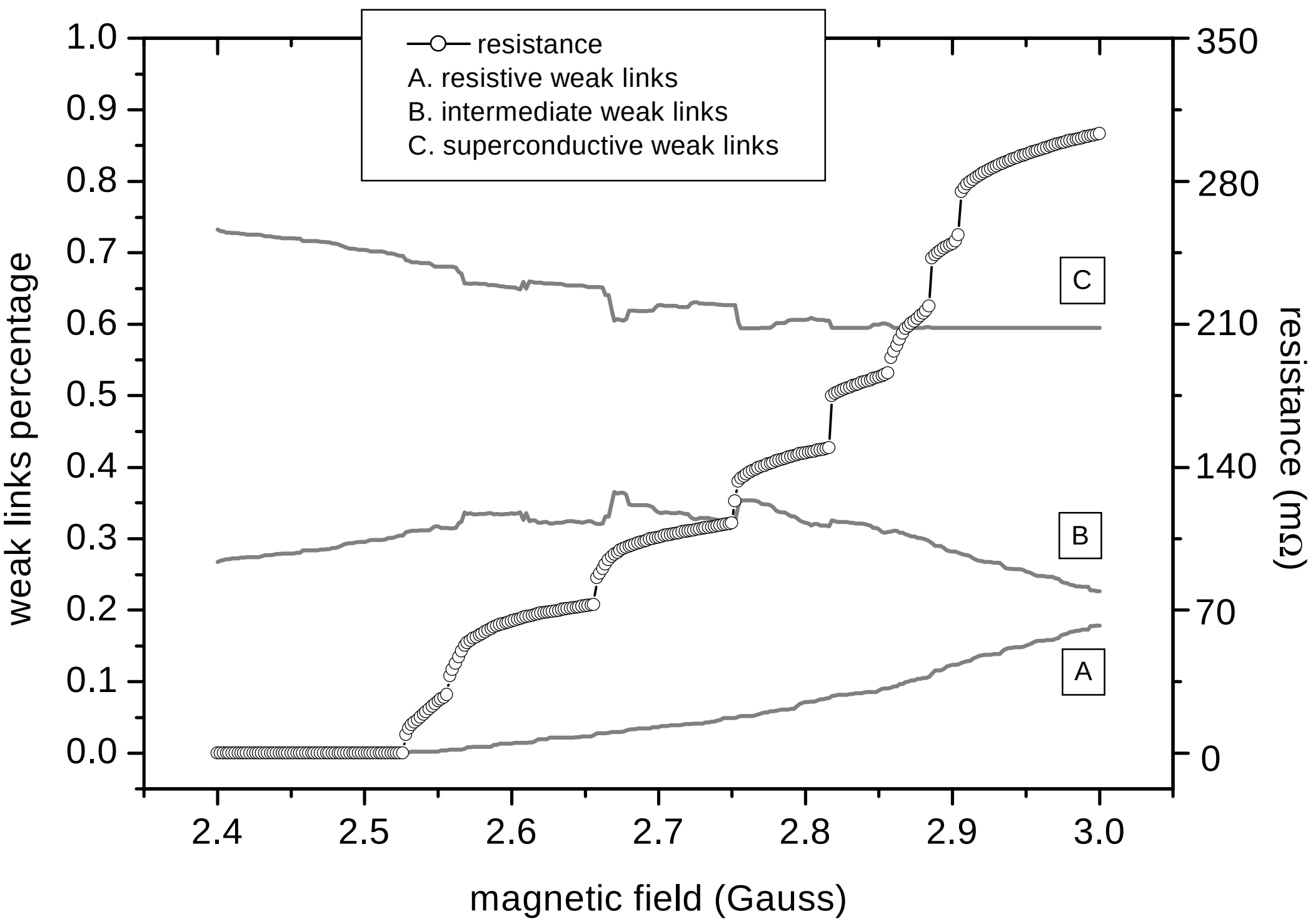

Figure 2: Circles represent the resistance increase versus magnetic field, obtained from a computer simulation, by considering the specimen as a 3-dimensional network of nonlinear resistors having Josephson-like V-I characteristics, representing the weak link ensemble. Curves A, B and C represent the field-dependent percentage of the weak links which are, respectively, in the resistive, intermediate and superconductive state during the transition process.

These results can be summarized as follows. When the current or the magnetic field are slowly changed, the specimen remains initially in a superconducting state until a critical value of the current or the field is reached. Above this critical value the specimen becomes resistive, and its resistance begins to increase continuously in a first stage, due to the increase of the number of weak links in an intermediate state (a state between superconducting and normal, where the differential resistance of a Josephson junction is very high). This stage of the process is apparently noise-free, but it is followed by a series of abrupt rearrangements of the spatial distribution of the superconducting, intermediate and normal state weak links, which correspond to a step-like increase of the specimen resistance (see Fig.2).
These rearrangements are due to the need of satisfy Kirchhoff equations for a network made up of highly non linear elements when the current or the field are changed, and are present even when the change is performed very smoothly.

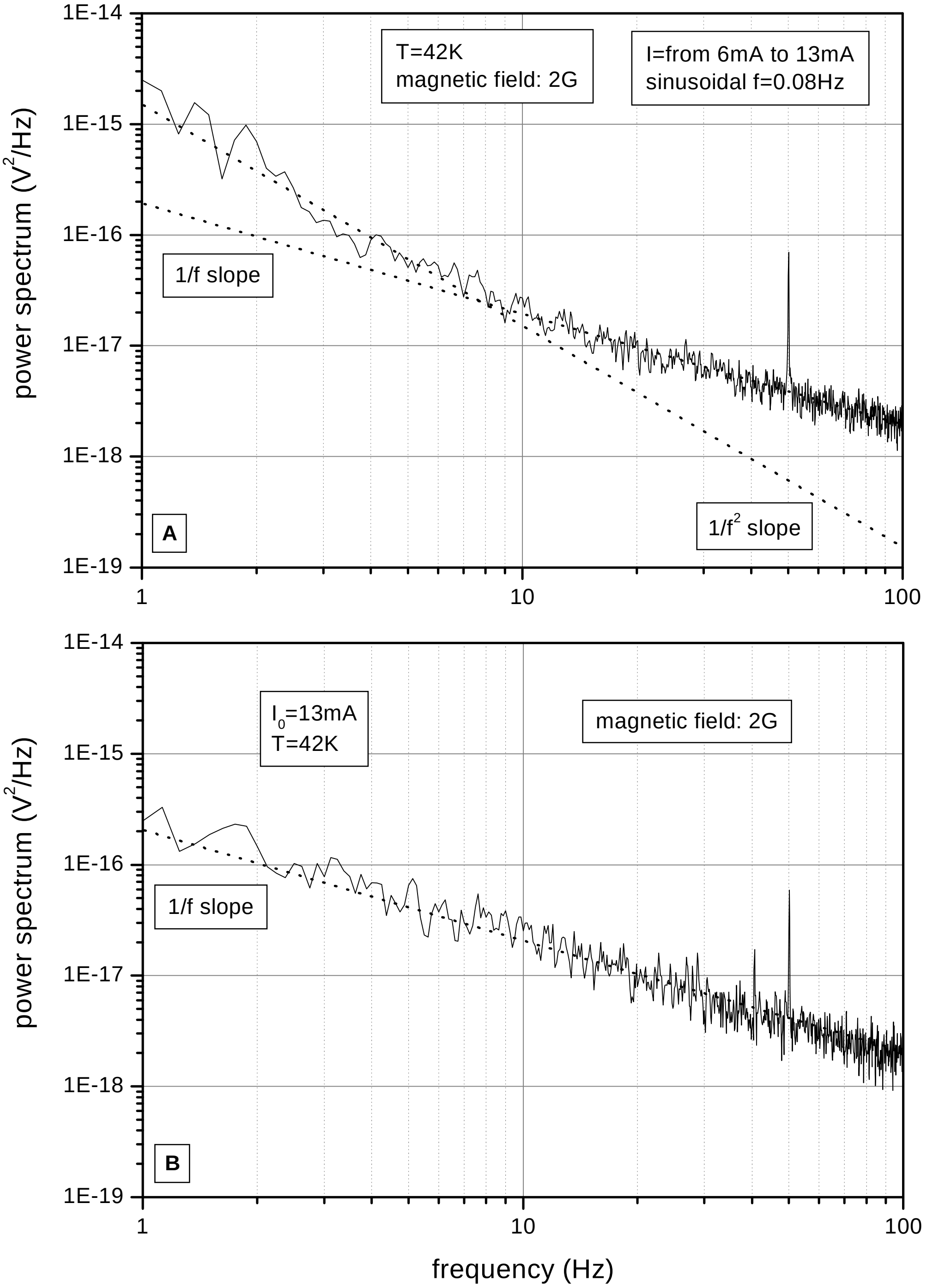

Figure 3: In Fig.3A power spectrum of the current noise in the presence of a sinusoidal a.c. current of 3.5 mA peak amplitude, having a frequency of 0.08 Hz . This a.c. current is superimposed to a d.c. current of 9.5 mA and gives, in the presence of a magnetic field of 2 G a resistance variation of 50 mΩ. The figure evidences the presence of a $1/f^2$ component in addition to a 1/f component, which is shown in Fig. 3B. In this second figure the noise spectrum is taken in stationary conditions, without the a.c. current and with a d.c. current ($I_0$) of 13 mA, corresponding to the maximum of the current flowing in the case of Fig.3A.

The presence of random steps in the resistance variation during cycling gives rise to a $1/f^2$ noise which can be experimentally detected (see Fig.3 and Fig.4). It can be observed that this noise component is superimposed to a 1/f component, which is present even in stationary conditions. Its origin is discussed in several publications, some of which are quoted in the previous paragraph.

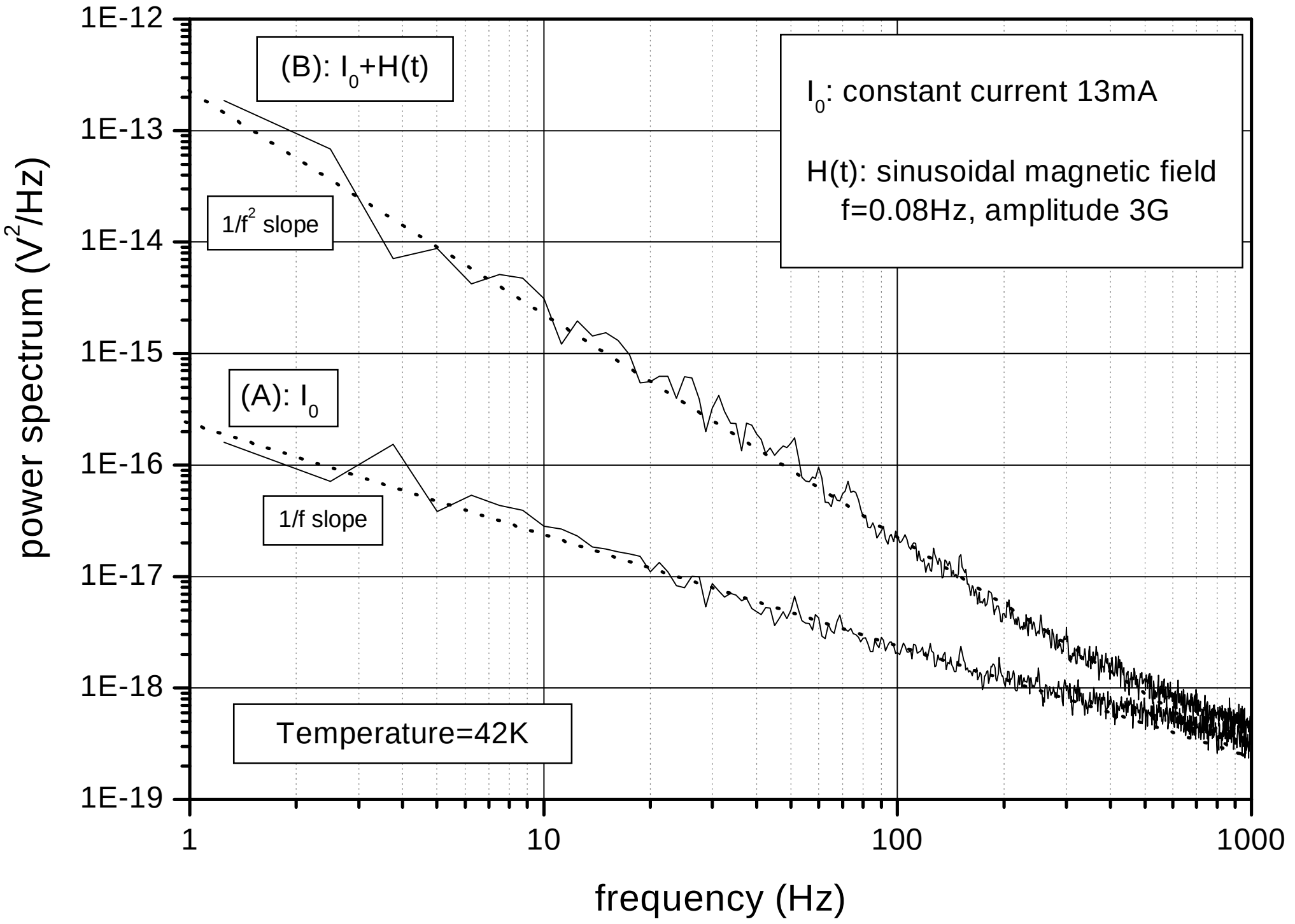

Figure 4: Power spectra of the current noise generated in the absence (curve A) and in the presence of an a.c. magnetic field (curve B). Spectrum A concerns the noise produced in stationary condition by a constant current $I_0$=13 mA, as in the case of Fig.3B, and is of the 1/f type. Spectrum B concerns the noise which is obtained by adding an a.c. magnetic field to the constant current. In this case it is a sinusoidal waveform of frequency 0.08 Hz and peak amplitude 3 G. It can be observed that the spectrum is of the $1/f^2$ type over nearly three decades of frequency and that its amplitude is much larger than the one reported in Fig. 3A, even if the resistance variation of the specimen during cycling was approximately the same.

While the model rightly accounts for the presence of a $1/f^2$ noise component during cycling, it cannot predict the amplitude of this component, since the computer simulation is performed over a network of a few hundreds nodes, several orders of magnitude too small to represent the actual specimen. On the other side, since Kirchhoff equations are solved iteratively over several hundreds of cycles and each curve requires an high density of points to evidence the steps, the computing time for larger networks becomes rapidly prohibitive.

Looking at the experimental results reported in Fig. 3 and Fig. 4 it may further be noticed that the $1/f^2$ noise component is much larger in the case where the magnetic field is cycled than in the case where the cycling concerns the current. In both cases the resistance variation during cycling is the same, so that it is not easy to explain this difference according to the model.

Actually the main difference between the two cases is that an increase of the magnetic field decreases the critical current of the weak links producing a certain amount of transitions, while an increase of the current does not change the critical current but has the same effect since it produces an equivalent amount of transitions. Even if the weak links involved in these transitions are not the same in the two cases, since the distribution of the field and of the current density within the specimen are completely different (the model takes into account the screening effect of the superconducting grains introducing a suitable distribution of the field over the field-sensitive elements constituting the 3-dimensional network), the very large increase of the $1/f^2$ noise component in the case of the field variation may not easily accounted for by the model.
Thus in this paper we propose an alternative interpretation of the origin of the current noise generated by a slowly changing magnetic field applied to a polycrystalline HTSC. It is based on the assumption that larger steps of resistance variation are produced by a discontinuous penetration of the field within the specimen, due to the presence of superconducting rings. These rings are constituted by different superconducting grains linked by weak links which are still in the superconducting state and form a closed circuit. Models based on superconductive rings were already proposed from several authors for explaining some properties of granular HTSC, but they have never been applied to explain the current noise produced by a varying magnetic field [9,10,11]. When the fields is varied, a current is induced in the rings which screens the field penetration, until the current exceeds the critical current of one or more weak links forming the ring and an abrupt variation of the field within the specimen takes place.
Once the field is penetrated within the broken ring, the current circulating in the ring vanishes, the ring is restored and a similar process takes place again. A simple calculation shows that, in the case of our specimen, the field which can be screened by the rings is very small with respect to the applied field, and thus this effect was disregarded in the development of the model described above. However the fact that after being broken the superconducting ring might be restored for a short time during the variation of the applied field was not properly taken into account, and this could actually be the cause for the observed discrepancy between current or field variation pointed out above.

## 3. CONCLUSION

In this paper we have presented some typical results concerning the current noise produced in a granular YBCO HTSC either by current or magnetic field variation. The noise is observed if the specimen is made resistive by the presence of a static magnetic field which disrupt a sufficiently large amount of weak links when the current is flowing, and is characterized by a $1/f^2$ spectrum. In addition to this noise component there is also a 1/f component which has a different origin, as pointed out in the previous section.
According to the model briefly described above, the $1/f^2$ spectral component is generated by a series of random steps of the electrical resistance during its increase or decrease by effect of a varying current or field, while the 1/f component may be generated by resistance fluctuations due to reversible transitions between superconductive and resistive state of those weak links which are near to a critical state, induced by temperature fluctuations [12,13,14,15]. In a different situation, when weak links do not play an important role, the noise may be predominantly generated by fluxoids motion. [4,5,16,17,18]. It is interesting to notice that series of random steps having the same sign when the resistance of the specimen increases or decreases gives rise in general to a $1/f^2$ component, while random steps of random signs may give rise to a 1/f spectrum under suitable conditions for their distribution. This second case is obviously the predominant one in stationary conditions.
In this paper we have furthermore evidenced that the $1/f^2$ component produced by a varying magnetic field might also have a different origin. Actually it may be generated by a discontinuous flux penetration within the specimen owing to the creation and disruption of superconducting rings involving several weak links. This might explain the large difference in the noise intensity produced by a varying field with respect to the one produced by a varying current.
In this context we can conclude that the model proposed in [7] might be rightly applied to explain the noise generated by current variation but that, as far as field variation is concerned, another more noisy process is the dominant one.